\documentclass[twocolumn,showpacs,preprintnumbers,prd,superscriptaddress,nofootinbib]{revtex4-1}
\usepackage{amsmath}
\usepackage{amsfonts}
\usepackage{amssymb}
\usepackage{graphicx}
\usepackage{color}
\usepackage{epstopdf}
\usepackage{hyperref}
\usepackage{booktabs}
\usepackage{hyperref}
\usepackage{cleveref}
\usepackage{color}
\Crefrangeformat{equation}{Eqs. (#3#1#4)--(#5#2#6)}
\Crefname{equation}{Eq.}{Eqs.}
\Crefname{figure}{Fig.}{Figs.}
\Crefname{section}{Sec.}{Secs.}

\begin{document}

\title{Model-independent reconstruction of $f(T)$ teleparallel cosmology}

\author{Salvatore Capozziello}
\email{capozziello@na.infn.it}
\affiliation{Dipartimento di Fisica, Universit\`a di Napoli  ``Federico II'', Via Cinthia, I-80126, Napoli, Italy.}
\affiliation{Istituto Nazionale di Fisica Nucleare (INFN), Sez. di Napoli, Via Cinthia 9, I-80126 Napoli, Italy.}
\affiliation{Gran Sasso Science Institute, Via F. Crispi 7, I-67100, L' Aquila, Italy.}

\author{Rocco D'Agostino} \email{rocco.dagostino@roma2.infn.it}
\affiliation{Dipartimento di Fisica, Universit\`a degli Studi di Roma ``Tor Vergata'', Via della Ricerca Scientifica 1, I-00133, Roma, Italy.}
\affiliation{Istituto Nazionale di Fisica Nucleare (INFN), Sez. di Roma ``Tor Vergata'', Via della Ricerca Scientifica 1, I-00133, Roma, Italy.}

\author{Orlando Luongo}	\email{luongo@na.infn.it}
\affiliation{Dipartimento di Fisica, Universit\`a di Napoli  ``Federico II'', Via Cinthia, I-80126, Napoli, Italy.}
\affiliation{Istituto Nazionale di Fisica Nucleare (INFN), Sez. di Napoli, Via Cinthia 9, I-80126 Napoli, Italy.}
\affiliation{School of Science and Technology, University of Camerino, I-62032, Camerino, Italy.}
\affiliation{Department of Mathematics and Applied Mathematics, University of Cape Town, Rondebosch 7701,
Cape Town, South Africa.}
\affiliation{Astrophysics, Cosmology and Gravity Centre (ACGC), University of Cape Town, Rondebosch 7701,
Cape Town, South Africa.}


\begin{abstract}
We propose a model-independent formalism to numerically solve the modified Friedmann equations in the framework of $f(T)$ teleparallel cosmology. Our strategy is to expand the Hubble parameter around the redshift $z=0$ up to a given order and to adopt cosmographic bounds as initial settings to determine the corresponding $f(z)\equiv f(T(H(z)))$ function. In this perspective, we distinguish two cases: the first expansion is up to the jerk parameter, the second expansion is up to the snap parameter. We show that inside the observed redshift domain $z\leq1$, only the net strength of $f(z)$ is modified passing from jerk to snap, whereas its functional behavior and shape turn out to be identical. As  first step, we set the cosmographic parameters by means of the most recent observations. Afterwards, we calibrate our numerical solutions with the concordance $\Lambda$CDM model. In both cases, there is a  good agreement with the cosmological standard model around $z\leq 1$, with severe discrepancies outer of this limit. We demonstrate that the effective dark energy term evolves following the test-function: $f(z)=\mathcal A+\mathcal Bz^2e^{\mathcal Cz}$. Bounds over the set $\left\{\mathcal A, \mathcal B, \mathcal C\right\}$ are also fixed by statistical considerations, comparing discrepancies between $f(z)$ with data. The approach opens the possibility to get a wide class of test-functions able to frame the dynamics of $f(T)$ without postulating any model \emph{a priori}. We thus re-obtain the $f(T)$ function through a back-scattering procedure once $f(z)$ is known. We figure out the properties of our $f(T)$ function at the level of background cosmology, to check the goodness of our numerical results. Finally, a comparison with previous cosmographic approaches is carried out giving  results  compatible with theoretical expectations.
\end{abstract}
\date{\today}

\maketitle


\section{Introduction}

A challenge of modern cosmology is to determine how the universe constituents precisely affect the dynamics as the universe expands. In particular, at late times of its evolution, the universe seems to be dominated by an exotic fluid dubbed dark energy, whose origin is so far unknown \cite{uno}. A wide number of observations points out that dark energy should act as a fluid with  negative pressure  counterbalancing the action of gravity, and speeding up the universe after a  transition epoch \cite{SNe,Planck15}. The standard approach which most likely aims to describe such a dynamics  assumes that the source of dark energy is supplied by a non-zero cosmological constant $\Lambda$, associated to vacuum quantum field fluctuations \cite{due}. The paradigm which makes use of $\Lambda$ is referred to as the $\Lambda$CDM concordance model and represents the simplest approach describing the observed universe dynamics \cite{tre}.

\noindent Although well-supported by observations, the concordance model does not give explanations towards the \emph{coincidence problem} \cite{Zlatev99} between dark matter and dark energy orders of magnitude and the \emph{fine-tuning} issue between predictions of quantum gravity and today observational constraints on  the value of  $\Lambda$  \cite{quattro}. Hence, a possible alternative interpretation is that the fluid which triggers the current universe speed up may not be due to $\Lambda$. For example, one can conclude that the effect of cosmic acceleration may be obtained in the framework of  modified gravities where extensions of  General Relativity (GR)  are accounted. In any gravitational theorie which extends   GR, additional degrees of freedom could be interpreted as dark energy sources \cite{cinque,francaviglia}.

\noindent In the context of modified gravity, it is possible to replace the Ricci scalar $R$ in the Einstein-Hilbert action by arbitrary functions of $R$ and any curvature invariant giving rise to Extended Theories of Gravity \cite{cinque,deFelice10,Odin}.
 However a torsional formulation of   GR, namely the \emph{Teleparallel Equivalent of   General Relativity } (TEGR) \cite{ft1} is also possible. In this approach, the gravitational action  is given  by the torsion scalar $T$  and  one can construct the $f(T)$ extension of TEGR \cite{storny1,Teleparallelism,Cai15}, where $f(T)$ is a function of   $T$. The predictions of   GR  turn out to coincide with TEGR, whereas the same does not happen for $f(T)$ with respect to 
 $f(R)$ gravity \cite{Cai15}. Indeed, $f(T)$ models lead to different outcomes if compared with the corresponding $f(R)$ approach. For these reasons, the $f(T)$ formulation of gravity represents an interesting approach   with a significative  number of cosmological implications which are today object of intensive investigation \cite{Teleparallelism,Cai15,storny2,storny3}.

In this paper we are going to investigate a   model-independent approach to reconstruct  the $f(T)$ function from cosmological observations. In this perspective, we consider the cosmolgical Friedmann equations for a generic  analytical  $f(T)$ function. Afterwards, we recast the corresponding equations in terms of the redshift and Hubble parameter $H(z)$. In so doing, we obtain a  differential equation which can be numerically solved, once $H(z)$ is somehow fixed by data. Even though we are interested in reconstructing $f(T)$, if we impose parametric forms of $H(z)$ we would fix the model \emph{a priori}, influencing our numerical analyses which turn out to be model dependent. To overcome this problem, we expand $H(z)$ in Taylor series around $z\simeq0$, which corresponds to present time $t_0$. Since cosmography or cosmo-kinetics lie on expansions of $a(t)$ around $t=t_0$, our expansions of $H$ can  be related to the cosmographic parameters, as in the case of $a(t)$. In particular, we consider two expansions: the former considering the cosmographic parameters $H_0,q_0$ and $j_0$, while the latter considering also  the snap parameter $s_0$. In such a picture, the cosmographic series does not require the assumption of any model \emph{a priori}. So that, using the most recent bounds over the cosmographic parameters, we are able to model-independently reconstruct $f(T)$. In fact, it is possible to  demonstrate that the above set of parameters univocally defines the shapes of $f(T)$ in the interval $0<z<2$. Furthermore, being $f(T)=f(H(z))$, it is possible to set a direct correspondence between $f(T)$ and $f(z)$. The former, $f(z)$, can be easily matched with arbitrary functions of the redshift $z$. Thus, we  first reconstruct $f(z)$ by means of the  cosmographic data  and afterwards we come back to $f(T)$ and its evolution in terms of $T$. 

In both  cases, we find a good agreement between the predicted and recovered shapes of $f(z)$ and $f(T)$ and is it possible to  notice that curve strengths are only modified, leaving unaltered the functional behaviors of each curves. To figure  out this result, we numerically analyze the obtained curves with a class of test-functions. We conclude that the best approach is offered by a particular class of combined exponential functions. With these considerations in mind, we obtain the corresponding $f(T)$ and  study the background cosmology. In particular, it is possible to describe the effective dark energy term and  its evolution in time. Finally, it is possible to fix limits over the equation of state (EoS) and  achieve  a form for $f(T)$ which well adapts to cosmic evolution at late times. 

The layout of the paper is the following. In \Cref{sezionedue} we briefly summarize the main ingredients of   $f(T)$ teleparallel gravity and   cosmology. In \Cref{sezionetre},  the basic requirements  to solve numerically the modified Friedmann equations are discussed. In particular,  the standard issues of cosmography are considered  and the main reasons  to conclude that cosmography is a model-independent approach to frame the universe dynamics are discussed. In \Cref{sezionequattro},  the method for numerically solving the Friedmann equations is implemented and   solutions for  $f(z)$ and $f(T)$  are introduced.  \Cref{sezionecinque} is devoted to the  investigation  of the obtained  $f(T)$ models and,  in \Cref{sezionesei}, we compare our results with previous ones obtained in  cosmography.  \Cref{sezionesette} is devoted to conclusions and perspectives.


\section{ $f(T)$ gravity and  cosmology}
\label{sezionedue}

The equations of motion in $f(T)$ gravities can be derived using as dynamical variables the veirbein fields $e_A(x^\mu)$, which form an orthonormal basis for the tangent space at each point $x^\mu$ of a generic manifold\footnote{Here, we use the capital Latin indices to denote the coordinate of the tangent space-time, while the Greek letters indicate the coordinates of the manifold.}. Introducing the dual basis $e^A(x^\mu)$, the metric tensor reads \cite{ft1}
\begin{equation}
g_{\mu\nu}=\eta_{AB}\ e^A_\mu e^B_\nu\,,
\end{equation}
where $\eta_{AB}=\text{diag}(1,-1,-1,-1)$ is the Minkowski metric of tangent space. The torsion tensor can be expressed by  the zero-curvature Weitzenb\"{o}ck connections $\hat{\Gamma}_{\mu\nu}^{\lambda}$ as \cite{Cai15}
\begin{equation}
T_{\mu\nu}^\lambda\equiv \hat{\Gamma}_{\mu\nu}^{\lambda}-\hat{\Gamma}_{\nu\mu}^{\lambda}=e^\lambda_A(\partial_\mu e_\nu^A-\partial_\nu e_\mu^A)\,.
\label{eq:torsion tensor}
\end{equation}
A convenient choice is to introduce the tensor
\begin{equation}
{S_{\rho}}^{\mu\nu}=\dfrac{1}{2}\left({K^{\mu\nu}}_\rho+\delta^\mu_\rho\ {T^{\alpha\nu}}_\alpha- \delta_\rho^\nu\ {T^{\alpha\mu}}_\alpha \right)\,,
\label{eq:superpotential}
\end{equation}
where
\begin{equation}
{K^{\mu\nu}}_\rho=-\dfrac{1}{2}\left({T^{\mu\nu}}_\rho-{T^{\nu\mu}}_\rho-{T_{\rho}}^{\mu\nu}\right)
\label{eq:cotorsion tensor}
\end{equation}
is the contorsion tensor. Hence, one can write the torsion scalar, which represents the teleparallel Lagrangian density, in the compact form
\begin{equation}
T={S_\rho}^{\mu\nu}{T^{\rho}}_{\mu\nu}\,.
\label{eq:torsion scalar}
\end{equation}
Generalizing to a generic function of $T$ means that the gravity action can be rewritten as:
\begin{equation}
S=\int d^4 x\ e \left[\dfrac{f(T)}{2\kappa}+\mathcal{L}_m \right]\,,
\label{eq:action}
\end{equation}
where $e=\sqrt{-g}=\det(e_\mu^A)$ and $\kappa=8\pi G$. We will assume throughout the paper the physical units with $\kappa=1$.  $\mathcal{L}_m$ is the Lagrangian density of the matter fields. The field equations are thus obtained by varying the action (\ref{eq:action}) with respect to the vierbein fields:
\begin{align}
&e_A^\rho{S_\rho}^{\mu\nu}(\partial_\mu T)f''+\left[\dfrac{1}{e}\partial_\mu(e e_A^\rho {S_\rho}^{\mu\nu})-e_A^\lambda {T^\rho}_{\mu\lambda}{S_\rho}^{\nu\mu}\right]f' \nonumber 	\\
&+\dfrac{1}{4}e_A^\nu f=\dfrac{\kappa}{2}e_A^\rho {{T^{(m)}}_\rho}^\nu
\end{align}
where ${{T^{(m)}}_\rho}^\nu$ represents the energy-momentum tensor of matter. In the above relations, the primes indicate derivatives with respect to $T$.

Hence, assuming the validity of the cosmological principle \cite{Peebles93}, we consider the homogeneous and isotropic Friedmann-Lema{\^i}tre-Robertson-Walker (FLRW) metric for a spatially flat universe\footnote{Generalizing  from spatially flat to non-flat is straightforward. However, spatial curvature would influence the cosmographic analysis, as we will discuss in \Cref{sezionetre}. In this paper we limit our attention to the flat case only.},

\begin{equation}
ds^2=dt^2-a^2(t)\left[dr^2+r^2(d\theta^2+\sin^2 \theta\ d\phi^2)\right]\ ,
\label{eq:metric}
\end{equation}

\noindent which corresponds to take the veirbein as: $e_A^\mu=\text{diag}(1,a,a,a)$. The modified Friedmann equations are thus \cite{sar}:
\begin{subequations}
\begin{align}
H^2&=\dfrac{1}{3}(\rho_m+\rho_T)\,,	\label{eq:Fried1}\\
\,\nonumber\\
2\dot{H}+3H^2&=-\dfrac{1}{3}(p_m+p_T)\,, \label{eq:Fried2}
\end{align}
\end{subequations}
where $\rho_m$ and $p_m$ represent the matter density and pressure for a perfect fluid source. In particular, we assume:
\begin{itemize}
  \item dark matter and dark energy do not interact between them;
  \item the matter density, $\rho_m$, scales as in standard cosmology, i.e. $\rho_m\propto a^{-3}$ with $a=(1+z)^{-1}$;
  \item the matter pressure is $p_m=0$ to guarantee a pressureless counterpart composed by baryons and cold dark matter;
  \item radiation, neutrinos, gravitational relics and so forth are negligibly small at our time and  cannot be considered in our analyses,
\end{itemize}
which turn out to be the basic demands of any cosmological theory to be consistent with current observations.
Moreover, in our picture $\rho_T$ and $p_T$ are the  torsion density and pressure respectively:
\begin{subequations}
\begin{align}
\rho_T&=Tf'(T)-\dfrac{f(T)}{2}-\dfrac{T}{2}\,, 	\label{eq:rho_T}\\
\,\nonumber\\
p_T&=\dfrac{f-Tf'(T)+2T^2f''(T)}{2[f'(T)+2Tf''(T)]}\,.\label{eq:p_T}
\end{align}
\end{subequations}
They are clearly zero for $f(T)=T$.
One can define an ``effective" dark energy torsional component, whose EoS takes the standard form in the hydrodynamic approach to cosmology  \cite{galaxy}. Since, for each species one has:
\begin{equation}\label{omegaspecie}
\omega_i\equiv \frac{p_i}{\rho_i}\,,
\end{equation}
and
\begin{equation}\label{omegatot}
\omega_{tot}\equiv \sum_i\left(\frac{ p_i}{\rho_i}\right)\,,
\end{equation}
then one gets:
\begin{equation}
\omega_{DE}=-1+\dfrac{(f-2Tf')(f'+2Tf''-1)}{(f+T-2Tf')(f'+2Tf'')}\,,
\label{eq:w_DE}
\end{equation}
where we assumed
\begin{equation}\label{assunzione}
\omega_{DE}\equiv \dfrac{p_T}{\rho_T}\,.
\end{equation}
It can be interpreted as  the torsional counterpart of $f(T)$ giving rise to  dark energy effect. Depending on the form of $f(T)$, it can  be the origin of today observed  acceleration of the Hubble flow. Using \Cref{eq:rho_T,eq:p_T}, the Friedmann equations can be rewritten as \cite{Aviles13}
\begin{subequations}
\begin{align}
H^2&=-\dfrac{1}{12f'(T)}\left[T\Omega_m+f(T)\right]\,, \label{eq:Friedmann1}  \\
\,\nonumber\\
\dot{H}&=\dfrac{1}{4f'(T)}\left[T\Omega_m-4H\dot{T}f''(T)\right]\,, \label{eq:Friedmann2}
\end{align}
\end{subequations}
where $\Omega_m$ is the normalized matter density parameter  whose functional form scales as: $\Omega_m\equiv\Omega_{m0}a^{-3}$, with $\Omega_{m0}$ the current value as measured by  observations. In the FLRW universe, the torsion scalar $T$ obeys the following constraint:
\begin{equation}
T=-6H^2\,,
\label{eq:T}
\end{equation}
which relates the torsion directly with the Hubble parameter at all stages of the universe evolution. With these considerations in mind, in the next paragraph, we  rewrite the Friedmann equations as function of the redshift  $z$ and $H$  finding the $f(T)$ form, once $H(z)$ is constrained by cosmography.


\section{ $f(T)$ cosmography}
\label{sezionetre}

Let us highlight now some basic assumptions of cosmography in view of  modeling the dynamics of $f(T)$  \cite{cosmografia1}. The expansion of  scale factor $a(t)$ in Taylor series around the present time $t_0$ \cite{cosmografia2} is 
\begin{equation}
a(t)=1+\sum_{k=1}^{\infty}\dfrac{1}{k!}\dfrac{d^k a}{dt^k}\bigg | _{t=t_0}(t-t_0)^k\,.
\label{eq:scale factor}
\end{equation}
From this expansion, one defines
\begin{align}
&H\equiv \dfrac{1}{a}\dfrac{da}{dt} \ , \hspace{1cm} q\equiv -\dfrac{1}{aH^2}\dfrac{d^2a}{dt^2}  \label{eq:H&q} \\
&j \equiv \dfrac{1}{aH^3}\dfrac{d^3a}{dt^3} \ , \hspace{0.5cm}  s\equiv\dfrac{1}{aH^4}\dfrac{d^4a}{dt^4}    \label{eq:j&s}
\end{align}
named \textit{Hubble}, \textit{deceleration}, \textit{jerk} and \textit{snap} parameters, respectively.
The luminosity distance can be written in terms of the redshift $z=a^{-1}-1$ as \cite{cosmografia3}
\begin{equation}
d_L(z)=(1+z)\int_0^z\dfrac{dz'}{H(z)}\,.
\label{eq:dL}
\end{equation}
We can expand in series \Cref{eq:dL} by simply plugging  \Cref{eq:scale factor} in it and then, by using the cosmographic parameters \Crefrange{eq:H&q}{eq:j&s} evaluated at our epoch, obtaining \cite{cosmografia4}:
\begin{equation}
d_L(z)=\dfrac{1}{H_0}\left(c_1 z+c_2 z^2 +c_3 z^3+c_4 z^4\right) + \mathcal{E}\,.
\label{eq:Taylor dL}
\end{equation}
Here, $\mathcal{E}$ is the error associated to the  truncation of  the series at a given order. In particular, indicating with $\delta_i$ the errors over the cosmographic coefficients, one gets:

\begin{eqnarray}
\left\{
  \begin{array}{ll}
    \mathcal E\rightarrow0, & \hbox{$N\rightarrow \infty$;} \\
    \delta\rightarrow\infty, & \hbox{$N\rightarrow \infty$.}
  \end{array}
\right.
\end{eqnarray}
with $N$ the order of truncating series. The consequences of such requests are that cosmography cannot be used with an arbitrary precision to frame the universe dynamics since any expansions is jeopardized by severe degeneracies among coefficients\footnote{For additional details see \cite{cosmografia4bis}}. This is not the only limit of cosmography but in this work we do not need to discriminate among models when $N$ significatively increases, but we only require cosmography to guarantee that $H$ is featured in a model-independent way \cite{cosmografia5}.
Thus, up to the fourth order, as reported in \Cref{eq:Taylor dL}, one gets:
\begin{align}
c_1& =1\ , \hspace{0.5cm} c_2=\dfrac{1}{2}(1-q_0)\ ,\nonumber \\
\,\nonumber\\
c_3&=-\dfrac{1}{6}\left(1-q_0-3q_0^2+j_0\right)\ , 		\label{eq:coeff dL}\\
\,\nonumber\\
c_4&=\dfrac{1}{24}\left(2-2q_0-15q_0^2-15q_0^3+5j_0+10q_0j_0+s_0\right)\ .\nonumber
\end{align}
Hence, using \Cref{eq:dL} one can relate the Hubble parameter to the luminosity distance by means of
\begin{equation}
H(z)=\left[\dfrac{d}{dz}\left(\dfrac{d_L(z)}{1+z}\right)\right]^{-1}\,,
\end{equation}
and so, making use of  \Cref{eq:Taylor dL}, one finds out the Taylor series expansion of $H$ as a function of the cosmographic parameters:

\begin{equation}
\frac{H(z)}{H_0}=1+\sum_{\tilde k=1}^{\infty}\dfrac{1}{\tilde k!}\dfrac{d^{\tilde k} H}{dz^{\tilde k}}\bigg | _{z=0}z^{\tilde k}\ ,
\label{eq:Taylor H}
\end{equation}
where $\tilde k\neq k$, while the first three orders read \cite{cosmografia6}
\begin{align}
\dfrac{d H}{dz}\bigg | _{z=0}& =1 + q_0\,,\nonumber\\
\dfrac{d^2 H}{dz^2}\bigg | _{z=0}&=j_0 - q_0^2\,,\\
\dfrac{d^3 H}{dz^3}\bigg | _{z=0}&=\dfrac{1}{6}\Big(j_0(3 +4q_0)-3q_0(1+q_0)+s_0\Big)\,.\nonumber
\end{align}
Now, if the values of the parameters $(H_0, q_0, j_0,s_0)$ are known,  one can combine \Cref{eq:Friedmann1,eq:Friedmann2} to numerically infer the function $f(T(z))=f(z)$. To figure  out this point, we need to convert time derivatives and derivatives with respect to $T$ into derivatives with respect to the redshift $z$. For any redshift-depending function $\mathfrak{F}(z)$, it is \cite{cosmografia7}:
\begin{align}
\dfrac{d}{dt}\mathfrak{F}(z)&=-(1+z)H(z)\dfrac{d}{dz}\mathfrak{F}(z) \ ,	\label{eq:dt-dz}	\\
\,\nonumber\\
\dfrac{\partial}{\partial T}\mathfrak{F}(z)&=-12H(z)H'(z)\dfrac{d}{dz}\mathfrak{F}(z) \label{eq:dT-dz}
\end{align}
where in the latter equation we have used \Cref{eq:T}. We will use \Cref{eq:dt-dz,eq:dT-dz} in what follows, in order to get the forms of the  functions $f(z)$ and  $f(T)$.


\section{Numerical reconstructions of $f(T)$ function}
\label{sezionequattro}

In this section we want to reconstruct the form of $f(z)$ by using the requirements imposed in the modified Friedmann equations. To do so, let us  consider \Cref{eq:Fried1,eq:dt-dz} which give the following differential equation in terms of $z$ and $H(z)$:
\begin{equation}
\left(\frac{df}{dz}\right)^{-1}\left[H (1+z) \frac{d^2f}{dz^2}+3 f \frac{dH}{dz}\right]=\mathcal D
\label{eq:f(z)}
\end{equation}
with
\begin{equation}\label{E}
\mathcal D \equiv \frac{1}{H}\left(\frac{dH}{dz}\right)^{-1}\left[3 \frac{dH}{dz}+(1+z) \frac{d^2H}{dz^2}\right]\,.
\end{equation}
As one can soon notice, \Cref{eq:f(z)} depends upon $z$ and $H(z)$. Assuming the form of $H(z)$  leads to impose the model and to force the consequent analysis. In other words, assuming a given $H(z)$, for example the one of the concordance $\Lambda$CDM model, would force $f(z)$ to be compatible with a slightly evolving dark energy term. Since we are looking for reconstructing the Hubble flow at different stages as model-independent as possible, we make use of cosmography to take $H$ free from any assumptions. Indeed, once the form of $H(z)$ is reconstructed by \Cref{eq:Taylor H}, we obtain a model-independent procedure which relates \Cref{eq:f(z)} to $z$ only, without passing through postulating the cosmological model. In particular, \Cref{eq:f(z)} is a second-order differential equation which requires two initial conditions over $f$ and $\frac{df}{dz}$. On the one hand, the initial condition on $\frac{df}{dz}$ can be obtained imposing that the present value of the effective gravitational constant is set to be the Newton constant $G$ \cite{cardone,Aviles13}:
\begin{equation}
\frac{df}{dz}\Big|_{z=0}=1 \ .
\label{eq:fp0}
\end{equation}
On the other hand, the initial condition over $f(z)$ comes from \Cref{eq:T,eq:fp0}:
\begin{equation}
f(T(z=0))=f(z=0)=6{H_0}^2(\Omega_{m0}-2)\,.
\label{eq:f0}
\end{equation}
Moreover, following the strategy in \cite{delaCruz16}, we may recast the cosmographic parameters $(q_0, j_0,s_0)$ as \cite{altro1}:
\begin{align}
&q_0=-1+\dfrac{3\tilde{\Omega}_{m0}}{2(1+2\tilde{F}_2)} \ ,	\label{eq:q0}\\
\,\nonumber\\
&j_0=1-\dfrac{9\tilde{\Omega}_{m0}^2(3\tilde{F}_2+2\tilde{F}_3)}{2(1+2\tilde{F}_2)^3} \ , \label{eq:j0} \\
\,\nonumber\\
&s_0=1-\dfrac{9\tilde{\Omega}_{m0}}{2(1+2\tilde{F}_2)}+\dfrac{45\tilde{\Omega}_{m0}^2(3\tilde{F}_2+2\tilde{F}_3)}{2(1+2\tilde{F}_2)^3} 	\label{eq:s0}\\
&\hspace{0.7cm}+\dfrac{27\tilde{\Omega}_{m0}^3(3\tilde{F}_2+12\tilde{F}_3+4\tilde{F}_4)}{4(1+2\tilde{F}_2)^4}-\dfrac{81\tilde{\Omega}_{m0}^3(3\tilde{F}_2+2\tilde{F}_3)^2}{2(1+2\tilde{F}_2)^5} 	\nonumber
\end{align}
with
\begin{equation}
\tilde{\Omega}_{m0}  =\dfrac{\Omega_{m0}}{F_1}\ , 	\hspace{0.5cm} \tilde{F}_i  =\dfrac{F_i}{F_1} \hspace{0.3cm} (i=2,3,4)
\end{equation}
\begin{equation}
F_i  =T_0^{i-1} f^{(i)}(T_0) \hspace{0.3cm}  (i=1,2,3,4) \ .
\end{equation}
With those assumptions in mind, we adopt the indicative values reported below \cite{delaCruz16}:
\begin{equation}
\left\{
\begin{aligned}
&\Omega_{m0}=0.289 \\
&h=0.692 \\
&q_0=-0.545\\
&j_0= 0.776\\
&s_0=-0.192
\end{aligned}
\right .
\label{eq:best-fit}
\end{equation}
where $h\equiv H_0/(100\ \text{km/s/Mpc})$. These results are compatible with the current expectations over the cosmographic set of parameters using different data sets \cite{altro2}.

\begin{figure}[h!]
\begin{center}
\includegraphics[width=3.2in]{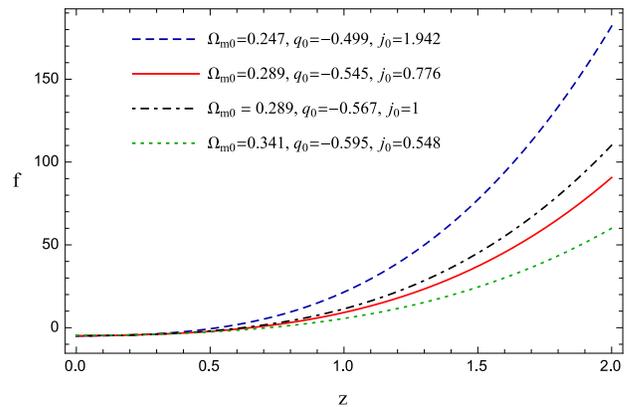}
\caption{Numerical reconstruction of $f(z)$ for different $(\Omega_{m0},q_0,j_0)$ based on the results of Table 6 in \cite{delaCruz16}. The solid red, the dashed blue and the dotted green lines correspond,  respectively,  to the the best-fit values, the upper $2\sigma$ bounds and the lower $2\sigma$ bounds of the parameters $(\tilde{\Omega}_{m0},\tilde{F}_2,\tilde{F}_3)$. The dot-dashed black line corresponds to the $\Lambda$CDM model. $h$ is fixed to the best-fit value of 0.692\ .}
\label{fig:f_z}
\end{center}
\end{figure}

Thus, we handle the aforementioned limits over the cosmographic parameters to numerically solve \Cref{eq:f(z)}. To do so, we perform a two-step  analysis. In particular, we first consider the Hubble rate expansion up to the second order in $z$, namely up to the jerk parameter. We then show our results in \Cref{fig:f_z} for different values of the set $(\Omega_{m0}, q_0,j_0)$, holding $H_0$ to be fixed.

We then match the numerical behaviours using the following test-functions\footnote{Here, test-functions are used to match \emph{a posteriori} the shapes of numerical curves obtained from our numerical analysis. The forms of such functions have been chosen analysing the shapes and checking any formal trends of the involved curves.}:
\begin{subequations}
\begin{align}
f_1(z)&=\mathcal A+z(\mathcal B+\mathcal C z)\ln(1+z^2)   \label{eq:ln}\\
f_2(z)&=\mathcal A+\mathcal B z^2e^{\mathcal C z}	           \label{eq:exp}\\
f_3(z)&=\mathcal A z^2+\mathcal B z\sin(1+\mathcal C z^2)	    \label{eq:sin}\\
f_4(z)&=\mathcal A+\mathcal B z^2\cos(1+\mathcal C z) 		       \label{eq:cos}\\
f_5(z)&=\mathcal A+\mathcal B\sinh(1+\mathcal C z)				\label{eq:sinh}\\
f_6(z)&=\mathcal A+\mathcal B z^3\tanh(\mathcal C z^2)		\label{eq:tanh}
\end{align}
\end{subequations}
with $ \mathcal A$, $\mathcal B$ and $\mathcal C$ free coefficients to be fixed from a direct comparison with the numerical curves.
To get suitable outcomes, we find a good analytical approximation according to the following requirements.
\begin{itemize}
  \item The function $f(z)$ has to be  neither odd nor even.
  \item The linear term, i.e. $f(z)\sim z$, is not favored and does not contribute significantly around the sphere $z\leq2$.
  \item Cosmography does not influence the functional analysis but rather it fixed the strengths of any test-function. The limits of cosmography are evident, since enlarging the prior domains leads to anomalous results on the form of $f(z)$ which are not compatible with current dark energy behaviour.
  \item The $f(z)$ results are all compatible with the concordance paradigm, but do not exclude dark energy to vary.
  \item The numerical analyses do not depend upon adding scalar curvature and radiation.
\end{itemize}
To find, among the test-functions, the best approximation for $f(z)$, we perform the $\mathcal{R}^2$-test \cite{Draper98}. Denoting by $f_i^{obs}$ the values of $f(z_i)$ obtained from the numerical solution, and by $f_i$ the correspondent values, we can define
\begin{equation}
\mathcal{R}^2\equiv1-\dfrac{\sum_{i=1}^n( f_i^{obs}-f_i)^2}{\sum_{i=1}^n (f_i^{obs}-\bar{f})^2}\ ,
\label{eq:R test}
\end{equation}
where
\begin{equation}
\bar{f}=\dfrac{1}{n}\sum_{i=1}^nf_i^{obs}
\end{equation}
and $n$ is the number of points. The $\mathcal{R}^2$ statistics provides information on how well the test-functions approximate the numerical $f(z)$, being $\mathcal{R}^2=1$ the ideal case in which the test-function agrees exactly with $f(z)$.

\begin{table}[h!]
\begin{center}
\begin{tabular}{|c|c|c|}
\hline
Test-function & $(\mathcal A, \mathcal B, \mathcal C)$ & $\mathcal{R}^2$  \\
\hline
$f_1(z)$ & $(-3.897,\ 10.88,\ 7.185)$ & 0.99974   \\
$f_2(z)$ & $(-5.024,\ 8.651,\ 0.512)$ & 0.99997   \\
$f_3(z)$ & $(15.73,\ -9.286,\ 1.112)$ & 0.99102   \\
$f_4(z)$ & $(-3.152,\ -21.52,\ 1.114)$ & 0.99630   \\
$f_5(z)$ & $(-10.93,\ 3.173,\ 1.593)$ & 0.99909   \\
$f_6(z)$ & $(-3.463,\ 11.89,\ 4.143)$ & 0.99959   \\
\hline
\end{tabular}
\caption{Results of the $\mathcal{R}^2$-test on the test-functions \Crefrange{eq:ln}{eq:tanh} using the parameters $(\Omega_{m0},h,q_0,j_0)=(0.289,\ 0.692,\ -0.545,\ 0.776)$ .}
\label{tab:test R^2}
\end{center}
\end{table}

\begin{figure}[!h]
\begin{center}
\includegraphics[width=3.1in]{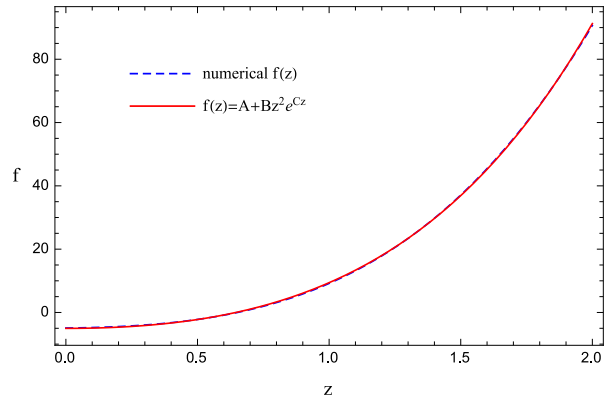}
\caption{Comparison between the numerical solution and the analytical approximation of $f(z)$. The parameters $(\Omega_{m0},h,q_0,j_0)$ are based on the best-fit results found in \cite{delaCruz16}, while the coefficients $( A,  B, C)$ are given in \Cref{eq:coeff 1}. }
\label{fig:num vs analy f}
\end{center}
\end{figure}

As an example, we consider  the red curve of \Cref{fig:f_z}. The $\mathcal{R}^2$-test suggests that the most suitable choice corresponds to the function (\ref{eq:exp}) (see \Cref{tab:test R^2}), i.e.
\begin{equation}
f(z)=\mathcal A+ \mathcal Bz^2e^{ \mathcal C z}
\label{eq:approx f}
\end{equation}
with
\begin{equation}
(\mathcal A, \mathcal B, \mathcal C)=(-5.024,\ 8.651,\ 0.512) \ .
\label{eq:coeff 1}
\end{equation}
However, it is evident from \Cref{tab:test R^2}  that very good approximations for $f(z)$ are also the functions \Cref{eq:ln} and \Cref{eq:tanh}, whose $\mathcal{R}^2$ values are far from the best one by only $0.024\%$ and $0.038\%$, respectively.
The comparison between the numerical solution of $f(z)$ and the functional form of \Cref{eq:approx f} is shown in \Cref{fig:num vs analy f} .

\begin{figure}[!h]
\begin{center}
\includegraphics[width=3.1in]{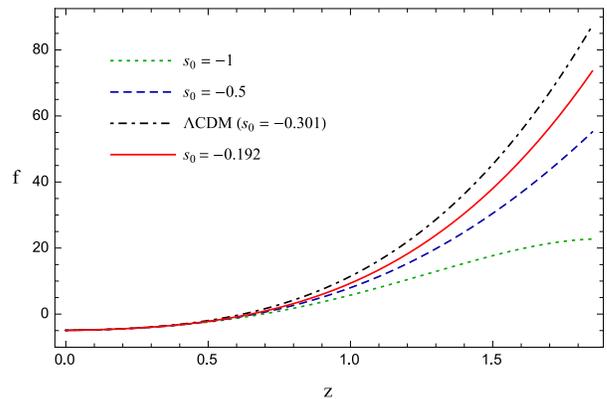}
\caption{Numerical solution of $f(z)$ for different values of the snap parameter. The parameters $(\Omega_{m0},h,q_0,j_0)$ are based on the best-fit results found in \cite{delaCruz16}. The solid red line corresponds to the best-fit value of $s_0$. The dot-dashed black line corresponds to the $\Lambda$CDM model with $\Omega_{m0}=0.289$. }
\label{fig:var s0}
\end{center}
\end{figure}

The second step takes $H(z)$ up to the third order, restricting the value of the snap parameter to the interval \cite{cosmosupp}
\vspace{-0.5cm}
\begin{equation}
s_0\in[-1,0] 	\ .
\label{eq:interval s0}
\end{equation}
In this case,  fixing  $(\Omega_{m0},h,q_0,j_0)$ to the best-fit values,  we show in \Cref{fig:var s0} the behaviour of $f(z)$ for different values of $s_0$. Consistently with what found before, the best approximation for $f(z)$ corresponding to the best-fit value of $s_0$ is provided by the same function as in \Cref{eq:approx f}, but with slightly different free parameters, namely:
\begin{equation}
(\mathcal A, \mathcal B, \mathcal C)=(-5.022,\ 8.577,\ 0.532)\ .
\label{eq:coeff 2}
\end{equation}
\vspace{-0.7cm}
\vspace{-0.7cm}

\newpage
\section{$f(T)$ cosmological models}
\label{sezionecinque}

\begin{figure}[h!]
\begin{center}
\includegraphics[width=3.2in]{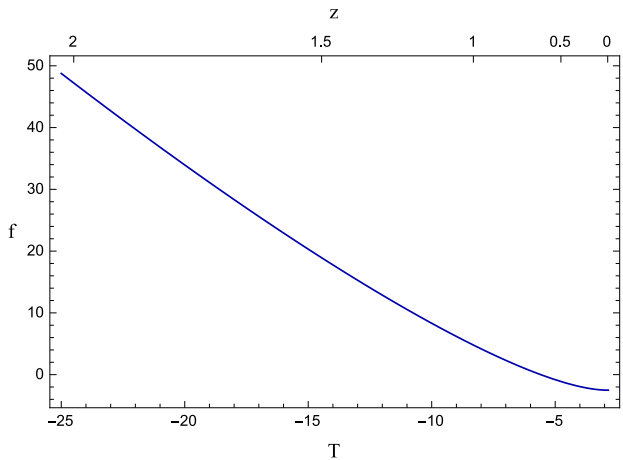}
\caption{Reconstructed $f(T)$ for the best-fit values of the cosmographic parameters and $\alpha=0.5$ in the redshift range $0\leq z \leq 2$. }
\label{fig:f(T)}
\end{center}
\end{figure}

\begin{figure}[!h]
\begin{center}
\includegraphics[width=3.2in]{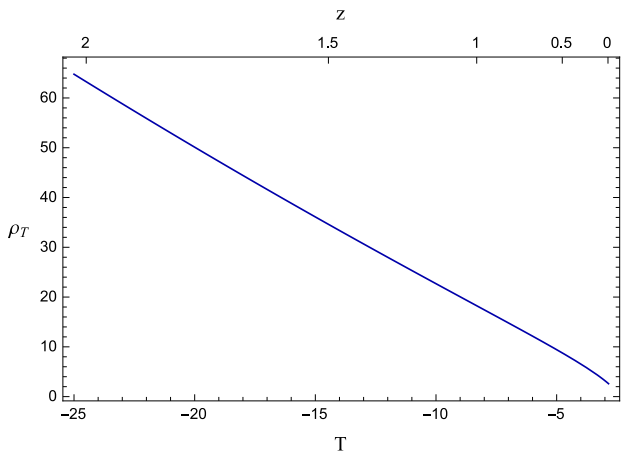}
\caption{Torsion density for the best-fit values of the cosmographic parameters and $\alpha=0.5$ in the redshift range $0\leq z\leq 2$.}
\label{fig:rho-T}
\end{center}
\end{figure}

\begin{figure}[!h]
\begin{center}
\includegraphics[width=3.2in]{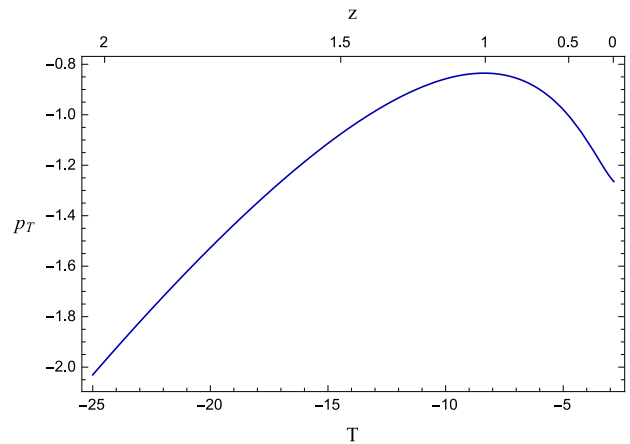}
\caption{Torsion pressure for the best-fit values of the cosmographic parameters in the redshift range $0\leq z\leq 2$. }
\label{fig:p-T}
\end{center}
\end{figure}

\begin{figure}[!h]
\begin{center}
\includegraphics[width=3.2in]{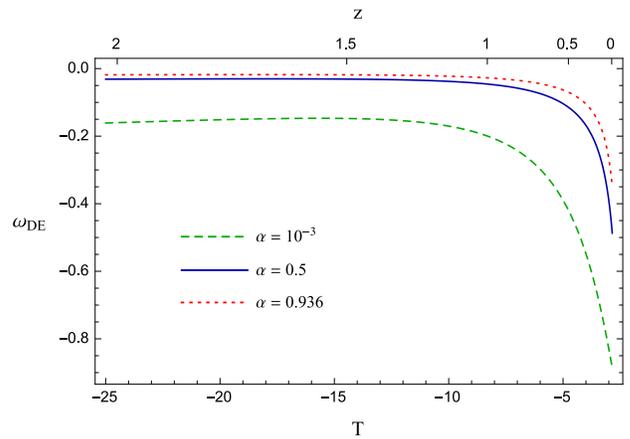}
\caption{Effective dark energy EoS parameter for the best-fit values of the cosmographic parameters in the redshift range $0\leq z\leq 2$. The different curves correspond to different values of the rescale factor $\alpha$ (cf. \Cref{eq:rescale factor}).}
\label{fig:w_DE-T}
\end{center}
\end{figure}

Our goal is now to reconstruct the function $f(T)$ through a back-scattering procedure. In particular, inverting the expansion of $H(z)$ up to the snap parameter by means of \Cref{eq:T}, one finds $z$ in terms of $T$, which can be inserted back into \Cref{eq:approx f} to finally reconstruct $f(T)$. To take into account the uncertainties in the estimate of the cosmographic parameters that may propagate through our numerical analysis, we rescale \Cref{eq:approx f} as
\begin{equation}
\alpha f(z)\rightarrow f(z)\ ,
\end{equation}
where the constant $\alpha$ will be determined from cosmological constraints. One obtains:
\begin{widetext}
\begin{equation}
z(T)=\dfrac{1}{2\mathcal{Q}}\left[2(q_0^2-j_0)+\left(\dfrac{4\mathcal{M}(T)}{H_0^3}\right)^{1/3}+\left(\dfrac{16H_0^3}{\mathcal{M}(T)}\right)^{1/3}\Big(j_0^2+q_0^2(6+12q_0+7q_0^2)-2j_0(3+7q_0+5q_0^2)-2s_0(1+q_0)\Big)\right]
\label{eq:z(T)}
\end{equation}
and
\begin{align}
f(T)=&\ \alpha \mathcal{A}+\dfrac{\alpha \mathcal{B}}{4\mathcal{Q}^2}\bigg[2(q_0^2-j_0)+\left(\dfrac{4\mathcal{M}(T)}{H_0^3}\right)^{1/3}+\left(\dfrac{16H_0^3}{\mathcal{M}(T)}\right)^{1/3}\Big(j_0^2+q_0^2(6+12q_0+7q_0^2)-2j_0(3+7q_0+5q_0^2) \nonumber \\
& -2s_0(1+q_0)\Big)\bigg]^2 e^{\frac{\mathcal{C}}{2\mathcal{Q}}\left[2(q_0^2-j_0)+\left(\frac{4\mathcal{M}(T)}{H_0^3}\right)^{1/3}+\left(\frac{16H_0^3}{\mathcal{M}(T)}\right)^{1/3}\Big(j_0^2+q_0^2(6+12q_0+7q_0^2)-2j_0(3+7q_0+5q_0^2)-2s_0(1+q_0)\Big)\right]}
\label{eq:f(T)}
\end{align}
where
\begin{equation}
\mathcal{M}(T)\equiv H_0^2\sqrt{2\mathcal{P}(T)}\mathcal{Q}-2H_0^3\mathcal{N}+\sqrt{-6T}H_0^2 \mathcal{Q}^2 \ ,
\label{eq:M}
\end{equation}
\begin{align}
\mathcal{P}(T)\equiv&\ 2H_0^2\bigg[6 j_0^3 - 6 q_0^2 \big(2 + q_0 (4 + q_0)\big)^2+3 j_0^2 \big(8 + q_0 (28 + 17 q_0)\big)+4  \Big(2 + q_0 \big(6 - q_0 (3 + 7 q_0)\big)\Big)s_0+9 s_0^2 \nonumber \\
&+2 j_0 \Big(6 (2 + 3 s_0) + q_0 \big(52 + q_0 (60 - q_0 (6 + 17 q_0)\big) + 27 s_0\Big)\bigg]-2 \sqrt{-6T} H_0 \Big(j_0^3+3 j_0^2 (1 + q_0) (6 + 11 q_0) \nonumber \\
&-3 j_0 q_0^2 \big(12 + q_0 (29 + 16 q_0)\big)+q_0^4 \big(18 + q_0 (36 + 17 q_0)\big)-15 q_0^2 (1 + q_0) s_0+3 j_0 (5 + 7 q_0) s_0 + 3 s_0^2\Big) \nonumber \\
&-3 T \big(-3 q_0^2 (1 + q_0) + j_0 (3 + 4 q_0) + s_0\big)^2\ ,
\label{eq:P}
\end{align}
\begin{equation}
\mathcal{Q}\equiv -3 q_0^2 (1 + q_0) + j_0 (3 + 4 q_0) + s_0 \ ,
\label{eq:Q}
\end{equation}
\begin{align}
\mathcal{N}\equiv &\ j_0^3 + 3 j_0^2 (1 + q_0) (6 + 11 q_0) - 3 j_0 q_0^2 \big(12 + q_0 (29 + 16 q_0)\big) + q_0^4 \big(18 + q_0 (36 + 17 q_0)\big) - 15 q_0^2 (1 + q_0) s_0  \nonumber \\
 &+3 j_0 (5 + 7 q_0) s_0 + 3 s_0^2\ .
\label{eq:N}
\end{align}
\end{widetext}
We can use the reconstructed $f(T)$ to study $\rho_T$ and $p_T$ and their cosmological implications and to compare our results with cosmological models developed in the literature so far \cite{h67}. So that, \Cref{eq:rho_T,eq:p_T} become:
\begin{widetext}
\begin{align}
\rho_T=&-\dfrac{1}{2}\left[T+\alpha\left(\mathcal{A}+\dfrac{\mathcal{B}\ \xi(T)}{4\mathcal{Q}^2}e^{\frac{\mathcal{C} \xi(T)}{2\mathcal{Q}}}\right)\right]+\dfrac{2^{1/3} \alpha\ \mathcal{B}\ T\ G(T) }{24H_0^3\mathcal{Q}^3\mathcal{M}(T)^2} X(T)\ Y(T)\ \mathcal{M}'(T)\ e^{\frac{\mathcal{C} \xi(T)}{2\mathcal{Q}}} \ ,
\label{eq:rho(T)}
\end{align}
\begin{align}
p_T=&\Bigg[e^{-\frac{\mathcal{C} \xi(T)}{2\mathcal{Q}}} H_0^4 Q^4 \mathcal{M}(T)^{10/3}\Bigg(72\mathcal{A}+\dfrac{18\mathcal{B} \xi(T)^2}{\mathcal{Q}^2}e^{\frac{\mathcal{C} \xi(T)}{2\mathcal{Q}}}-\dfrac{3\times 2^{1/3}\mathcal{B} T G(T) X(T) Y(T) \mathcal{M}'(T)}{H_0^3\mathcal{Q}^3\mathcal{M}(T)^2}e^{\frac{\mathcal{C} \xi(T)}{2\mathcal{Q}}}+\dfrac{2^{1/3}\mathcal{B} T^2 }{H_0^3\mathcal{Q}^4\mathcal{M}(T)^{10/3}}  \nonumber \\
&\times e^{\frac{\mathcal{C} \xi(T)}{2\mathcal{Q}}} \bigg(2^{1/3}\mathcal{C} G(T)^2  X(T) Y(T)\mathcal{M}'(T)^2+2^{7/3}H_0\mathcal{Q} X(T) Y(T) \mathcal{M}(T) \mathcal{M}'(T)^2 +2H_0\mathcal{Q} G(T) \mathcal{M}(T)^{1/3}	 \nonumber \\
& \times \Big(-6 X(T) Y(T) \mathcal{M}'(T)^2+2^{5/3} \big(\mathcal{C} X(T)+Y(T)\big)\mathcal{M}(T)^{2/3} \mathcal{M}'(T)^2 +2H_0\big((2\mathcal{Q}+\mathcal{C}(q_0^2-j_0))X(T)\ \nonumber \\
& +(q_0^2-j_0)Y(T)\big) \mathcal{M}(T)^{1/3}\mathcal{M}'(T)^2 + 3X(T) Y(T) \mathcal{M}(T) \mathcal{M}''(T)\Big)\bigg)\Bigg)\Bigg]\times \Bigg[2^{4/3}\mathcal{B}\bigg(2^{1/3}\mathcal{C} G(T)^2  X(T) Y(T)\mathcal{M}'(T)^2 \nonumber \\
&+2^{7/3}H_0\mathcal{Q} T  X(T) Y(T)	 \mathcal{M}(T) \mathcal{M}'(T)^2+H_0\mathcal{Q} G(T) \mathcal{M}(T)^{1/3}\Big(-12 T X(T) Y(T) \mathcal{M}'(T)^2+2^{7/3} \big(\mathcal{C} X(T) \nonumber \\
& +Y(T)\big) T  \mathcal{M}(T)^{2/3} \mathcal{M}'(T)^2+4H_0\big((2\mathcal{Q}+\mathcal{C}(q_0^2-j_0))X(T)+(q_0^2-j_0)Y(T)\big)T \mathcal{M}(T)^{1/3}\mathcal{M}'(T)^{2} \nonumber \\
&+ 3X(T) Y(T) \mathcal{M}(T)\big(\mathcal{M}'(T)+2 T \mathcal{M}''(T)\big)\Big)\bigg)\Bigg]
\label{eq:p(T)}
\end{align}
where
\begin{equation}
\xi(T)\equiv 2(q_0^2-j_0)+\left(\frac{4\mathcal{M}(T)}{H_0^3}\right)^{1/3}+\left(\frac{16H_0^3}{\mathcal{M}(T)}\right)^{1/3}\ ,
\label{eq:xi}
\end{equation}
\begin{equation}
X(T)\equiv 2^{4/3}H_0^2\mu+2H_0(q_0^2-j_0)\mathcal{M}(T)^{1/3}+2^{2/3}\mathcal{M}(T)^{2/3}\ ,
\label{eq:X}
\end{equation}
\begin{equation}
Y(T)\equiv 2^{4/3}\mathcal{C} H_0^2\mu+2H_0\left(2\mathcal{Q}+\mathcal{C}(q_0^2-j_0)\right)\mathcal{M}(T)^{1/3}+2^{2/3}\mathcal{C} \mathcal{M}(T)^{2/3}\ ,
\end{equation}
\begin{equation}
G(T)\equiv -2H_0^2\mu+2^{1/3}\mathcal{M}(T)^{2/3}
\label{eq:G}
\end{equation}
\begin{equation}
\mu\equiv  j_0^2+q_0^2(6+12q_0+7q_0^2)-2j_0(3+7q_0+5q_0^2) -2s_0(1+q_0)
\label{eq:mu}
\end{equation}
and
\begin{equation}
\mathcal{M}'(T)\equiv\dfrac{\partial \mathcal{M}(T)}{\partial T}=\dfrac{1}{\sqrt{2}}\left[\dfrac{\sqrt{3}H_0^2\mathcal{Q}^2 T}{(-T)^{3/2}}+\dfrac{H_0^2\mathcal{Q}}{\sqrt{\mathcal{P}(T)}}\mathcal{P}'(T)\right]\ ,
\label{eq:Mp}
\end{equation}
\begin{equation}
\mathcal{M}''(T)\equiv\dfrac{\partial^2 \mathcal{M}(T)}{\partial T^2}= -\dfrac{\sqrt{-3T}H_0^2\mathcal{Q}^2\mathcal{P}(T)^2+H_0^2\mathcal{Q}T^2 \mathcal {P}(T)^{1/2}\mathcal{P}'(T)^2-2H_0^2\mathcal{Q}T^2\mathcal{P}(T)^{3/2}\mathcal{P}''(T)}{2\sqrt{2}T^2\mathcal{P}(T)^2}    \ ,
\label{eq:Mpp}
\end{equation}
\begin{align}
\mathcal{P}'(T) \equiv &\ \dfrac{\partial \mathcal{P}(T)}{\partial T}= -3 \left(-3 q_0^2 (1 + q_0) + j_0 (3 + 4 q_0) + s_0\right)^2+\dfrac{\sqrt{6}H_0}{\sqrt{-T}}\Big[j_0^3 + 3 j_0^2 (1 + q_0) (6 + 11 q_0) - 3 j_0 q_0^2 \big(12 \nonumber \\
& + q_0 (29 + 16 q_0)+q_0^4 \big(18 + q_0 (36 + 17 q_0)-15 q_0^2 (1 + q_0) s_0\big)+3 j_0 (5 + 7 q_0) s_0 + 3 s_0^2\Big]\ ,
\label{eq:Pp}
\end{align}
\begin{align}
\mathcal{P}''(T)\equiv &\ \dfrac{\partial^2 \mathcal{P}(T)}{\partial T^2}=\sqrt{\dfrac{3}{2}}\dfrac{H_0}{(-T)^{3/2}}\Big[j_0^3 + 3 j_0^2 (1 + q_0) (6 + 11 q_0)-3 j_0 q_0^2 \big(12 + q_0 (29 + 16 q_0)\big)+q_0^4 \big(18 + q_0 (36 + 17 q_0)\big) \nonumber \\
& -15 q_0^2 (1 + q_0) s_0 + 3 j_0 (5 + 7 q_0) s_0 + 3 s_0^2\Big]\ .
\label{eq:Ppp}
\end{align}
\vspace{-0.5cm}
\end{widetext}
We notice that $p_T$ is in fact independent of the rescaling factor $\alpha$. From the condition of having an accelerating expansion today, we are able to constrain $\alpha$ by imposing that $\omega_{DE}<-1/3$. Thus,
\begin{equation}
0<\alpha \lesssim 0.936\ .
\label{eq:rescale factor}
\end{equation}
\Cref{fig:f(T)} shows the reconstructed $f(T)$ for the best-fit values of the cosmographic parameters with the indicative value of $\alpha=0.5$.
The physical density $\rho_T$ and pressure $p_T$ are shown in \Cref{fig:rho-T,fig:p-T} , respectively, while we show in \Cref{fig:w_DE-T} the effective dark energy EoS parameter as defined in \Cref{eq:w_DE}.
\vspace{-0.2cm}
\vspace{-0.5cm}
\vspace{-0.5cm}

\section{Comparing  $f(T)$ models }
\label{sezionesei}

\begin{figure}[!h]
\begin{center}
\includegraphics[width=3.2in]{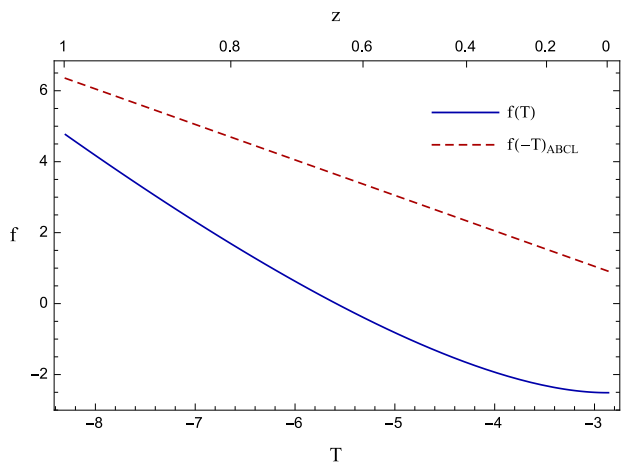}
\caption{Comparison between $f(T)$ found in this work for $\alpha=0.5$ (solid blue line) and  $f(|T|)$ proposed in \citep{Aviles13} (dashed red line) in the redshift interval $0\leq z \leq 1$. }
\label{fig:comp f(T)}
\end{center}
\end{figure}

In this section, we compare the reconstructed $f(T)$ model with the one proposed in \cite{Aviles13}, which has been obtained by means of cosmography. Both the approaches might be in agreement to guarantee the goodness of our numerical analysis, in the redshift domain in which cosmography is valid. So that, let us recall the model defined in \cite{Aviles13}
\begin{align}
f(T)_\text{ABCL}&=	c_0T+(T-T_0)\Big[c_1+c_3\cosh(T-T_0) +\nonumber \\
&(T-T_0)\Big(c_2+c_4(T-T_0)\sinh(T-T_0)\Big)\Big]
\label{eq:ABCL f(T)}
\end{align}
where $T_0=-6H_0^2$. The model is composed by distinct parts, each of them dominates over the other as the redshift increases. In particular, we choose the free parameters $c_i\  (i=0,\hdots,4)$ imposing that $f(T_0)=6H_0^2(\Omega_{m0}-2)$ and $f'(T_0)=1$, as well as we made as initial settings of our approach. So that, by using the derivative constraints on $f(T)$ at $z=0$, obtained from a direct experimental analysis with different data sets, one gets:
\begin{subequations}
\begin{align}
&c_0=2-\Omega_{m0}\\
&c_1=\Omega_{m0}-1\\
&c_2=-3\times 10^{-6}\\
&c_3=\dfrac{1}{15}\times 10^{-9}\\
&c_4=\dfrac{3}{4}\times 10^{-14}
\end{align}
\end{subequations}
while the best-fit values of the present matter density parameter and the Hubble constant were found to be, respectively
\begin{equation}
\Omega_{m0}=0.364\ , \hspace{0.5cm}  H_0=71.47\ \text{km/s/Mpc} \ .
\end{equation}
Without taking into account the sign of $T$, let us consider the comparison between our present model and \Cref{eq:ABCL f(T)}. It is clear, from \Cref{fig:comp f(T)}, that the two models are compatible for $z\leq 1$. Discrepancies have been accounted since the previous model has been obtained by means of heuristic results due to cosmography, whereas our model is a numerical solution of the modified Friedmann equations. Discrepancies are around the $10\%\div15\%$ and testify the goodness between the two approaches.
We conclude that both the models are similar to the concordance paradigm and seem to slightly depart from it at the observable limit of $z>1$. More significant deviations are however available as one exceeds such an interval, i.e. for $z>1$. The limit of cosmography, found in the model \eqref{eq:ABCL f(T)} is overcome by extending it with the numerical analysis performed in the present work.
\vspace{-0.2cm}
\vspace{-0.5cm}
\section{Conclusions}
\label{sezionesette}

We considered  $f(T)$ teleparallel cosmological  models. In particular, we proposed a new model-independent strategy to reconstruct the $f(T)$ function, without {\it a priori}  assumptions over the model. To figure out this,  we recast the Hubble parameter $H(z)$ in terms of cosmographic parameters and  expanded it up to the jerk first and snap later. We noticed that our outcomes have not been influenced by the order, leaving unaltered the functional behaviors of $f(T)$. So, since numerical bounds over the cosmographic series can be  found from kinematics only, without requiring any other assumptions, we rewrote the modified Friedmann equations in terms of the redshift only and  recast  $f(T)$ as function of the redshift $z$ through the constraint $T=-6H^2$, with $H=H(z)$, previously expanded in Taylor series. In particular, we set the cosmographic parameters by means of the most recent observations first, and, later, with the concordance $\Lambda$CDM predictions. 

Our treatment enabled us to get a single set of differential equations all in terms of the redshift, which can be numerically solved in order to get $f(z)\equiv f(T(H))$. So that, using the most recent bounds over the cosmographic coefficients, we inferred $f(z)$ through test-functions. All the adopted test functions passed the cosmographic initial settings and framed the universe dynamics with a precise accuracy. We handled several test-functions and  found that the best agreement can  be accounted if  $f(z)=\mathcal A+ \mathcal Bz^2e^{ \mathcal C z}$. In this perspective, we considered a statistical $\mathcal{R}^2$-test, to get the significance between our model and the proposed test-functions. We thus found the set of free parameters to be $(\mathcal A, \mathcal B, \mathcal C)=(-5.022,\ 8.577,\ 0.532)$ and  noticed that $f(z)$ is neither odd nor even, being constructed without linear terms $\propto z$. Afterwards, we came back to $f(T)$ and its evolution in terms of $T$, with a backscattering procedure, i.e. inverting $z$ in terms of $T$, by means of $T=-6H^2$. Therefore, we investigated some consequences of the obtained $f(T)$ model and  wrote its equation of state, which turns out to be different from the concordance case, providing a slightly varying effective dark energy term. Finally, we compared our results with the ones proposed in previous cosmographic $f(T)$ models. We showed that, at least in the redshift domain $z\in[0,1]$, the approach that made use of cosmography agrees with our prescription. Outer this interval, however, the effective dark energy term turns out to be consistently different from a pure cosmological constant, providing that the $\Lambda$CDM model can be seen as a limit of a more general paradigm included in $f(T)$ gravity cosmologies. Future works will include refined numerical tests in view to extend  the numerical behaviors of $f(z)$ toward early phase cosmology at higher $z$.

\section*{Acknowledgments}
S.C. acknowledges the support of INFN (iniziativa specifica QGSKY). This paper is based upon work from COST action CA15117 (CANTATA), supported by COST (European Cooperation in Science and Technology).

\end{document}